\documentclass[showpacs,amsmath,amssymb,twocolumn,aps]{revtex4}

\input{epsf}
\begin{document}
\title{Photogalvanic current in electron gas over a liquid helium surface}

\author{M.V. Entin$^{1}$, L.I. Magarill$^{1,2}$}
\affiliation{$^{1}$Institute of Semiconductor Physics, Siberian
Branch of the Russian Academy of Sciences, Novosibirsk, 630090, Russia
\\ $^{2}$ Novosibirsk State University, Novosibirsk, 630090, Russia}
\begin{abstract}
We study the  stationary surface photocurrent in 2D electron gas near the helium surface. Electron gas is assumed to be attracted to the helium surface due to the image attracting force and an external stationary electric field.  The alternating electric field has both vertical and in-plane components. The photogalvanic effect originates from the periodic transitions of electrons between quantum subbands in
the vertical direction caused by a normal component of the
alternating electric field accompanied by synchronous in-plane
acceleration/deceleration due to the  electric
field in-plane component. The effect needs vertical asymmetry of the system. The problem is considered
 taking into account a friction caused by the electron-ripplon interaction. The photocurrent   resonantly  depends on the field frequency. The resonance occurs at field frequencies close to the distance between  well subbands. The resonance is symmetric or antisymmetric depending on the kind (linear or circular) of polarization.
\end{abstract}

   \maketitle

\subsection*{Introduction}
The surface photogalvanic effect (PGE) (the stationary  in-plane photocurrent) arises in confined systems.  This photocurrent exists even if crystal asymmetry is negligible,
  but the quantum well is oriented (up and down normals are not equivalent).
  The current along the surface occurs if the microwave electric field  has both in- and
  out-plane components. Different solid systems have been examined,  classical
  \cite{we4} and quantum \cite{we5,taras1} films,  and the systems with a single boundary where the  in-plane PGE current flows   in the vicinity of this boundary \cite{alper,gus,we6}.

The phenomenology of surface PGE in the absence of magnetic field
is determined by the relation for the current density
\begin{equation}\label{phenomen}
   {\bf j}=\alpha^s\Big(({\bf E}-{\bf n}({\bf nE}))({\bf n
   E^*})+c.c.\Big)+i\alpha^a[{\bf n}[{\bf E}{\bf E^*}]],
\end{equation}
where ${\bf n}$ is the normal to the quantum well, ${\bf
E}(t)=\mbox{Re}({\bf E}e^{-i\omega t})$ is the uniform microwave
electric field. Real constants $\alpha^s$ and $\alpha^a$
describe linear and circular photogalvanic effects,
correspondingly. The origin of this current can be understood if
to consider the out-of-plane electric field component as
modulating the quantum well conductivity with a simultaneous
driving of electrons by the in-plane field. More recent papers by the authors deal with PGE in the classical parabolic potential well   with inhomogeneous vertical distribution of impurities \cite{we_jetpl1} and the double quantum well \cite{we_jetpl2}.

Recently, this effect has been experimentally studied in relation to the electron gas  over the helium surface \cite{alex_chep} in  the presence of a magnetic field.
The electron gas over the liquid helium surface (EGLH) has remained a popular 2D system since 1970th, when the first papers about this system appeared (see, e.g., \cite{shik_mon}). The advantage of this system is the possibility to realize the conducting medium with a very low electron concentration, $<10^6$cm${}^{-2}$, which is much lower than that in a solid state system. The absence of impurity scattering provides a very large electron mobility as compared to solid systems. At the same time, EGLH possesses the electron-ripplon scattering mechanism that  differs EGLH from  solid systems.

The purpose of the present paper is the theoretical study of the PGE in quantum gas over the helium surface without magnetic field. The system under consideration is depicted in Fig.1. Electrons are attracted to the liquid helium via electrostatic polarization, but they can not enter inside helium due to the barrier. The polarization attraction of electrons to liquid helium leads to the appearance of 2D electron subbands. Thermal electrons occupy the bottom of the lower subband. In a quantum well the vertical component of alternating electric field  can cause the transitions  between different quantum
subbands. In the presence of scattering the alternating electric field gives birth to the
effective pumping of the in-plane momentum to the electronic
system. The microwave field  plays the role of the energy and asymmetry
source, while the scatterers produce  in-plane acceleration of
electrons.

The indirect photoexcitation is a two-stage process with the participation of the intermediate state. The resonance behavior of the indirect transition probability  manifests itself when the photon energy approaches to the distance between subbands. Parallelism of 2D subbands leads to the independence of this resonance from the electron momentum and, hence, to the similar resonance in the overall stationary current.
\begin{figure}[h]\label{fig1}
\centerline{\epsfxsize=8cm\epsfbox{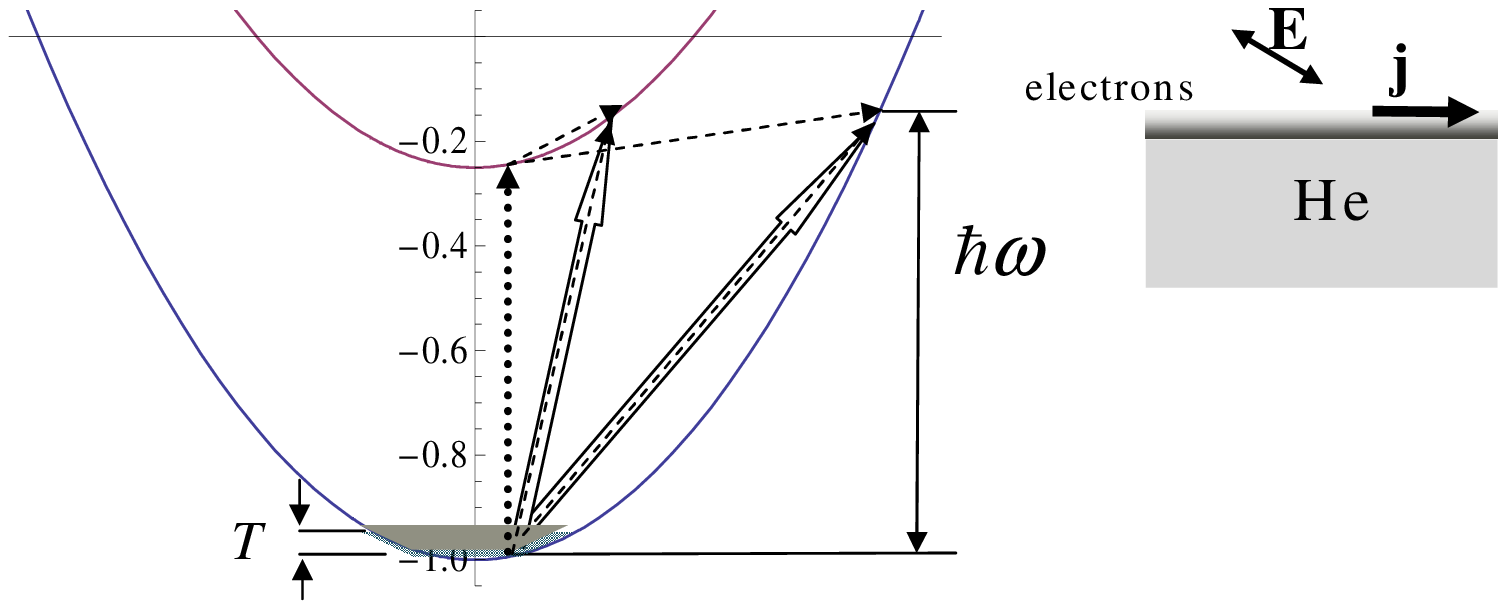}}
 \caption{Left panel: The sketch of transitions. Equilibrium electrons occupy the temperature layer at the bottom of the first subband. Electrons experience indirect phototransitions from the first subband to empty states of the first or second subband (empty arrows) with the participation of ripplons. The amplitudes of ripplon and optical transitions are depicted by dashed and dotted line, correspondingly. The excitation is a result of interference of optical (vertical dotted lines) and ripplon (tilted dashed lines) transition amplitudes. Right panel: A sketch of the proposed experiment. Tilted alternating electric field causes the stationary current in the electron gas over the helium surface.}
\end{figure}

 To understand the effect in more details, let us discuss the problem  of classical electrons confined by a   well and affected by a tilted uniform weak alternating electric field and the friction force strongly decreasing with the distance to the surface \cite{we_jetpl1} (this is the case of the electron-ripplon interaction which will be assumed below). Well  self-frequency $\omega_0$ corresponds to the intersubband distance of the quantum case.

An electron vibrates or circulates in response to the external electric field. If the external field is weak and circular-polarized, the rotation occurs in out-of-resonance with $\omega_0$   and in the exact resonance when the field is linear polarized. The inhomogeneous friction converts the rotation into a progressive motion, while it does not do that with the vibration. Hence, the stationary photocurrent in the linear-polarized microwave field  should occur in exact resonance conditions while the response to the circular polarization changes its sign near the resonance frequency.

\subsection*{The problem formulation}
We study electrons confined by the image forces near the helium surface.
The electron states are described by a 2D momentum along the surface,  ${\bf p}=(p_x,p_y)$, and the discreet quantum number $n$ corresponding to the motion in z-direction. The transversal quantized states of electron $\chi_n(z)$  are  hydrogen-like. Energies $\epsilon_{n,{\bf p}}$ and wave functions $\psi_{n,{\bf p}}({\bf r},z)$ of electron states  are
\begin{eqnarray}\label{epsilon}
\epsilon_{n,{\bf p}}=\frac{p^2}{2m}+\varepsilon_n,~~ \varepsilon_n=-\varepsilon_B/n^2,\\ \psi_{n,{\bf p}}({\bf r},z)=\frac{1}{\sqrt{S}}\chi_n(z)\exp(i{\bf pr}); \\ \chi_1(z)=\frac{2z}{a_B^{3/2}}e^{-z/a_B}, \nonumber \\
\chi_2(z)=\frac{z}{\sqrt{2a_B}a_B}(1-z/2a_B)e^{-z/2a_B},
\end{eqnarray}
where $\varepsilon_B=1/2ma_B^2$ and $a_B=\kappa/me^2$ are the effective Bohr energy and Bohr radius, $m$ is the electron mass,  $\kappa=4\kappa_1(\kappa_1+\kappa_2)/(\kappa_2-\kappa_1)$, $\kappa_1$, $\kappa_2$ are the dielectric constants of gaseous and liquid helium, correspondingly, $S$ is the system area  (we set $\hbar=1$).
The non-diagonal matrix element of the transversal coordinate $z$, which will be required below, is $z_{12}=-32a_B\sqrt{2}/81$.

We will consider the photogalvanic effect at the frequency close to the intersubband distance $\Delta=\varepsilon_2-\varepsilon_1=3/8ma_B^2=3\varepsilon_B/4$. In this case only states 1 and 2 are actual.

The estimates show that the main scattering mechanism is the electron-ripplon scattering. The ripplons are  surface-tension-controlled vibrations of the helium surface \cite{shik_mon}. In our case, the energy of emitted/absorbed ripplons is much lower than the electron energy. Really, let us consider the kinematics of the ripplon emission/absorption process: $\epsilon_{n,{\bf p}}=\epsilon_{n',{\bf p}'}\pm \omega_{\bf q}$. A typical ripplon wave vector has the order of the thermal electron wave vector. Then $\omega_q/T \ll 1$.
As a result, the process is quasi-elastic. At electron density $10^6\mbox{cm}^{-2}$ characteristic for the experiment with liquid helium \cite{alex_chep}   and temperature $T\sim 0.1$K, the electron gas is non-degenerate and the ripplon energy is much less than $T$. In this case the ripplons produce static fluctuation potential. That means possibility to use approach of \cite{we_jetpl2} valid for the impurity scattering.

Assuming that the mean free time is large, as compared to the distance between the levels of quantum wells (and also temperature), one can treat $n$ and ${\bf p}$ as good quantum numbers and describe the problem within the kinetic equation for distribution functions $f_{n,{\bf p}}$. In such an equation,  external classical alternating electric field ${\bf E}(t)$ causes the transition between unpertubed states and determines the generation term in the kinetic equation.

The kinetic equation for the first stationary correction $f^{(1)}_{n,{\bf p}}$ to
equilibrium distribution function $f^{(0)}_{n,{\bf p}}$  reads
\begin{equation}\label{kin_eq}
    \sum_{n',{\bf p}'}(W_{n',{\bf p}';n,{\bf p}}f^{(1)}_{n',{\bf p}'}-W_{n,{\bf p};n',{\bf p}'}f^{(1)}_{n,{\bf
    p}})+ G_{n,{\bf p}} =0.
    \end{equation}
The first term in  kinetic equation (\ref{kin_eq}) corresponds to the relaxation due to the electron-ripplon scattering with transition probability $W_{n,{\bf p};n',{\bf p}'}$.  The generation due to a combined action of the external electric field and the scattering is represented by term $G_{n,{\bf p}}$. Just this combined action causes the pumping of in-plane momentum to the system, which is necessary for the in-plain current. This term is quadratic in the electric field.

Note that the  classical kinetic equation (\ref{kin_eq}) neglects the
off-diagonal elements of the density matrix. This assumption is valid if the subbands collision broadening is less than the distance between them.

Generation $G_{n,{\bf p}}$ is given by
\begin{equation} \label{G}
   G_{n,{\bf p}}= \sum_{n',{\bf p}'}w_{n,{\bf p};n',{\bf p}'}(f^{(0)}_{n',{\bf p}'}-f^{(0)}_{n,{\bf
    p}}),
\end{equation}
where $w_{n,{\bf p};n',{\bf p}'}$ is the transition probability with the accounting for the microwave electric field and electron-ripplon interaction,  $f^{(0)}_{n,{\bf  p}}$ is the equilibrium distribution function.  Then $w_{n,{\bf p};n',{\bf p}'}$ is determined by the second order perturbation term which  includes the Hamiltonians of electron interaction with
electromagnetic field $H_{ph}$ and electron-ripplon interaction
$H_{e-r}$.   Operator $H_{ph}$ is
\begin{equation}\label{Hph}
    H^{ph}=-\frac{e}{\omega}\mbox{Im}\left({\bf E} e^{-i\omega t}\right){\bf v}\equiv \frac{1}{2}(Ue^{-i\omega t} + h.c.),
\end{equation}
where  ${\bf v}=({\bf p}/m,v^z)$ is
the velocity operator.  Operator $U $ can be presented in the form: $U= ie({\bf p}{\bf E}_\|/m+v_zE_z)/\omega$.

  Free ripplons are described by  the Hamiltonian
 \begin{equation}\label{Hr} H_r=\sum_{\bf q} \omega_{\bf q}b^+_{\bf q}b_{\bf q},\end{equation}
  where $b^+_{\bf q}$, $b_{\bf q}$ are the ripplon creation and destruction operators. The interaction of a single  electron with ripplons  is given by (see \cite{shik_mon})
  \begin{equation}\label{Her}
   H_{er}=S^{-1/2}\sum_{\bf q} (b^+_{-\bf q}+b_{\bf q})e^{i\bf qr}V_{\bf q}(z),
\end{equation}
where in our case of prevailing image forces and infinite helium layer (see, e.g., \cite{shik_mon})
\begin{equation}\label{Vq}
   V_q(z)=\frac{1}{ma_B^3}\sqrt{\frac{q}{2\rho\omega_q}}\bar{V}_q(z), ~~~~ ~\bar{V}_q(z)=\frac{a_B^2}{z^2}(1-qzK_1(qz)).
\end{equation}

Expressions describing the resonance PGE obtained  in \cite{we_jetpl2} can be adapted for the case of electron-ripplon interaction considered here.
Taking the quasielasticity of the electron-ripplon scattering into account,  the probability satisfies the relation of reversibility $W_{n',{\bf p}';n,{\bf p}}=W_{n,{\bf p};n',{\bf p}'}$. In this case we get
\begin{equation}\label{W}
    W_{n,{\bf p};n',{\bf p}'} = \frac{2\pi }{S} (2N_{|{\bf p'-p}|}+1)((V_{|{\bf p'-p}|})_{n,n'})^2\delta(\epsilon_{n,{\bf p}}-\epsilon_{n',{\bf p}'}),
\end{equation}
where $~(V_q)_{nn'}= \int_0^\infty dz \chi_n(z)\chi_{n'}(z)V_q(z), ~~N_q=1/(\exp{(\omega_q/T)}-1)$ is the Bose-Einstein distribution function,  $ \omega_q = \sqrt{\sigma_0q^3/\rho}$ is the frequency of ripplon (gravitational part of the ripplon dispersion is neglected), $\rho$ is the liquid helium density, $\sigma_0$ is the  helium surface tension coefficient.

The excitation probability
including the electron-ripplon scattering is determined by the
second-order transition amplitude. The needed term arises from the
interference of amplitudes caused by the $E_z$ and ${\bf E}_{||}$. The draft of the transitions is
depicted in  Fig.1.

In the second interaction order,  for the transition
probability, one can obtain
\begin{eqnarray}\label{Wph}\nonumber
&&w_{n,{\bf p};n',{\bf p}'}= \frac{\pi}{2S}(2N_{|{\bf p'-p}|}+1)\times \Bigg[\delta(\epsilon_{n,{\bf p}}-\epsilon_{n',{\bf p}'}+\omega)\times  \nonumber \\ && \Bigg|\sum_{n_1}\Bigg(\frac{(V_{|{\bf p-p'}|})_{n,n_1}U^+_{n_1,{\bf
p}';n',{\bf p}'}}{\eta
i(\varepsilon_{n_1,n'}+\omega)}+ \nonumber \\ && \frac{U^+_{n,{\bf
p};n_1,{\bf p}}(V_{|{\bf p-p'}|})_{n_1,n'}}{\eta +
i(\varepsilon_{n_1,n}-\omega)}\Bigg)\Bigg|^2 +  \delta(\epsilon_{n,{\bf p}}-\epsilon_{n',{\bf p}'}-\omega)\times \nonumber \\ && \Bigg|\sum_{n_1}\Bigg(\frac{(V_{|{\bf p-p'}|})_{n,n_1}U_{n_1,{\bf p}';n',{\bf p}'}}{\eta +
i(\varepsilon_{n_1,n'}-\omega)}+\nonumber \\   && \frac{U_{n,{\bf
p};n_1,{\bf p}}(V_{|{\bf p-p'}|})_{n_1,n'}}{\eta +
i(\varepsilon_{n_1,n}+\omega)}\Bigg)
\Bigg|^2\Bigg];~~~~~~~~~~~~~~(\eta = +0).
\end{eqnarray}
Here $\varepsilon_{n,n'}\equiv \varepsilon_{n}-\varepsilon_{n'}$.
The denominators in Eq.(\ref{Wph}) have their  resonance with the field frequency
independently from the  electron momentum. At the same time, the resonance
in the final state is absent due to non-conservation of the
in-plane momentum.

The stationary current density is given by
\begin{equation}\label{j}
    {\bf j}=\frac{2e}{S}\sum_{n,{\bf p}}\frac{\bf p}{m}f^{(1)}_{n,{\bf
    p}}.
\end{equation}
\subsection*{Photocurrent}
According to Eq. (\ref{j})  the PGE current is determined by the  first angular harmonic of the distribution function $f^{(1)}_{n,{\bf
    p}}$.
    The system of kinetic equation (\ref{kin_eq}) can be presented in  the  algebraic form:
 \begin{eqnarray} \label{kin_eq1}
    \frac{1}{\tau_n(\epsilon)} {\bf  A}_n(\epsilon)-\frac{1}{\tau_{n,\bar{n}}(\epsilon)} {\bf A}_{\bar{n}}(\epsilon)={\bf B}_n(\epsilon),
 \end{eqnarray}
      where
      \begin{eqnarray}\label{AB}
   {\bf A}_n(\epsilon)=\Big\langle\frac{2{\bf p}}{p^2}f^{(1)}_{n,{\bf
    p}}\Big\rangle,~~
    {\bf B}_n(\epsilon)=\Big\langle\frac{2{\bf p}}{p^2}G_{n,{\bf p}}\Big\rangle.\end{eqnarray}
    Here $$\Big\langle\ldots\Big\rangle = \frac{1}{2\pi}\int_0^{2\pi}d\phi(\ldots),$$
   $\phi$ is the electron momentum angle, $p=\sqrt{2m(\epsilon-\varepsilon_n)}$, a line bar over $n$   means: $\bar{n} =2$ if $n=1$ and vice versa.

Using Eq.(\ref{W}) one can find for the relaxation times introduced in  Eq.(\ref{kin_eq1}):
\begin{eqnarray}\label{nuintra}\nonumber
   && \frac{1}{\tau_n(\epsilon)}= \frac{4\pi T}{S} \Big\langle\sum_{\bf  q}\frac{1}{\omega_q}\Big[(V_{q})_{nn}^2\delta(\epsilon-\epsilon_{n,{\bf p+q}})\times \nonumber \\ &&\Big(-\frac{{\bf qp}}{p^2}\Big)+
    (V_{q})_{n\bar{n}}^2\delta(\epsilon-\epsilon_{\bar{n},{\bf p+q}})\Big]\Big\rangle,\\
        \label{nuinter} &&\frac{1}{\tau_{n,\bar{n}}(\epsilon)} =
         \frac{4\pi T}{S} \Big\langle\sum_{\bf q}\frac{1}{\omega_q}
   (V_{q})_{n\bar{n}}^2\delta(\epsilon-\epsilon_{\bar{n},{\bf p+q}})\times \nonumber \\ &&\Big(1+\frac{{\bf qp}}{p^2}\Big)\Big\rangle.
    \end{eqnarray}

The PGE current is determined by the  odd part of  transition probability $w_{n ,{\bf p};n',{\bf p}'}$ (denoted by tilde):
\begin{eqnarray}\label{wpart}
    &&\tilde{w}_{n,{\bf p};n,{\bf p}'}= \nonumber\\&& \frac{2\pi Te^2}{S m\omega^3} \mbox{Re}\Bigg
[\frac{(V_{|{\bf p'-p}|})_{n,n}(V_{|{\bf p'-p}|})_{n,\bar{n}}}{\omega_{|{\bf p'-p}|}}\times\nonumber \\ && \Bigg((({\bf p'-p}){\bf E}_\|^*)\Bigg(\frac{v^z_{n,\bar{n}}E_z}{i\eta +
(\varepsilon_{\bar{n},n}+\omega)} + \frac{v^z_{\bar{n},n}E_z}{i\eta +
(\varepsilon_{\bar{n},n}-\omega)}\Bigg)
\times\nonumber\\&&\delta(\epsilon_{n,{\bf p}}-\epsilon_{n,{\bf p}'}+\omega)
 \nonumber -\\&&(({\bf p'-p}){\bf E}_\|)\Bigg(\frac{v^z_{n,\bar{n}}E_z^*}{i\eta +
(\varepsilon_{\bar{n},n}-\omega)} + \frac{v^z_{\bar{n},n}E_z^*}{i\eta +
(\varepsilon_{\bar{n},n}+\omega)}\Bigg)\nonumber \times\\&& \delta(\epsilon_{n,{\bf p}}-\epsilon_{n,{\bf p}'}-\omega)\Bigg)
\Bigg]; \end{eqnarray}
\begin{eqnarray} \label{wpart1}
&&\tilde{w}_{n,{\bf p};\bar{n},{\bf p}'}= \nonumber\\ && \frac{2\pi Te^2}{S m\omega^3} \mbox{Re}\Bigg
[\frac{(V_{|{\bf p'-p}|})_{n,\bar{n}}}{\omega_{|{\bf p'-p}|}}\times \nonumber \\
&&\Bigg((({\bf p'-p}){\bf E}_\|^*)\Bigg(\frac{(V_{|{\bf p'-p}|})_{n,n}v^z_{\bar{n},n}E_z}{i\eta +(\varepsilon_{n,\bar{n}}+\omega)} + \nonumber \\
&&\frac{(V_{|{\bf p'-p}|})_{\bar{n},\bar{n}}v^z_{\bar{n},n}E_z}{i\eta +
(\varepsilon_{\bar{n},n}-\omega)}\Bigg)
\delta(\epsilon_{n,{\bf p}}-\epsilon_{\bar{n},{\bf p}'}+\omega)
 \nonumber - \\
&& (({\bf p'-p}){\bf E}_\|)\Bigg(\frac{(V_{|{\bf p'-p}|})_{n,n}v^z_{\bar{n},n}E_z^*}{i\eta +
(\varepsilon_{n,\bar{n}}-\omega)} + \nonumber \\
&&\frac{(V_{|{\bf p'-p}|})_{\bar{n},\bar{n}}v^z_{\bar{n},n}E_z^*}{i\eta +
(\varepsilon_{\bar{n},n}+\omega)}\Bigg)\delta(\epsilon_{n,{\bf p}}-\epsilon_{\bar{n},{\bf p}'}-\omega)\Bigg)
\Bigg].
   \end{eqnarray}

From Eq.(\ref{j}) it follows:
\begin{equation}\label{j1}
    {\bf j}=\frac{e}{mS}\sum_{n,{\bf p}}{\bf A}_n(\epsilon_{n,p})p^2=\frac{em}{\pi}\int_{\varepsilon_n}^\infty (\epsilon-\varepsilon_n){\bf A}_n(\epsilon) d\epsilon.
    \end{equation}
Solving the system of Eq.(\ref{kin_eq1}) we find for ${\bf A}_n$ at $\epsilon>\varepsilon_2$:
\begin{eqnarray}
\label{An}
  {\bf A}_n(\epsilon)=  \frac{\tau_n(\epsilon)[{\bf B}_n(\epsilon)+\tau_{\bar{n}}(\epsilon)\tau_{n,\bar{n}}(\epsilon)^{-1}{\bf B}_{\bar{n}}(\epsilon)]}
  {1-\tau_1(\epsilon)\tau_2(\epsilon)\tau_{1,2}(\epsilon)^{-1}\tau_{2,1}(\epsilon)^{-1}}. \end{eqnarray}
In the region $\varepsilon_1<\epsilon<\varepsilon_2$
\begin{equation}\label{An0}
   {\bf A}_2(\epsilon)=0, ~~~~ {\bf A}_1(\epsilon)=\tau_1(\epsilon){\bf B}_1(\epsilon).
\end{equation}
Using  Eq.(\ref{AB}) and Eqs.(\ref{G},\ref{wpart},\ref{wpart1})  we obtain in the resonance ($\omega \simeq \Delta$) approximation:
  \begin{eqnarray} \label{Bn}
  {\bf B}_n(\epsilon)=2c~\mbox{Re}({\bf E}_\|^*E_z)f_s(\delta)s_n(\epsilon)+ 2c~\mbox{Im}({\bf E}_\|^*E_z)f_a(\delta)a_n(\epsilon),\nonumber\end{eqnarray}
 \begin{eqnarray}f_s(\delta)=\frac{\eta}{\eta^2+\delta^2}~~~~f_a(\delta)=\frac{\delta}{\eta^2+\delta^2}\end{eqnarray}

 \begin{eqnarray}\label{sn} && {s}_n(\epsilon)=\frac{9\pi^2 \sigma_0\Delta }{16 m\omega^3S}e^{-(\epsilon-\varepsilon_1)/T}\Big\langle\sum_{\bf q}\frac{({\bf qp})(V_q)_{1,2}}{p^2\omega_q}\times \nonumber \\
 && \Bigg[(V_q)_{n,n}\Big(\delta(\epsilon-\epsilon_{n,{\bf p+q}}+\omega)(e^{-\omega/T}-1)  +\nonumber \\
&&\delta(\epsilon-\epsilon_{n,{\bf p+q}}-\omega)(e^{\omega/T}-1)\Big) - \nonumber \\ && (-1)^n\Big((V_q)_{1,1}+(V_q)_{2,2}\Big)\delta(\epsilon_{n,p}-\epsilon_{\bar{n},{\bf p+q}}-(-1)^n\omega)\times \nonumber \\
&&(e^{(-1)^n\omega/T}-1)
\Bigg]\Big\rangle,\\
 \label{an}&& {a}_n(\epsilon)=\frac{9\pi^2 \sigma_0\Delta }{16 m\omega^3S}(-1)^ne^{-(\epsilon-\varepsilon_1)/T}\Big\langle\sum_{\bf q}\frac{({\bf qp})(V_q)_{1,2}}{p^2\omega_q}\times \nonumber \\
 && \Bigg[(V_q)_{n,n}\Big(\delta(\epsilon-\epsilon_{n,{\bf p+q}}+\omega)(e^{-\omega/T}-1)  -\nonumber \\
&&\delta(\epsilon-\epsilon_{n,{\bf p+q}}-\omega)(e^{\omega/T}-1)\Big) - (e^{(-1)^n\omega/T}-1)\times\nonumber \\ && \Big((V_q)_{1,1}-(V_q)_{2,2}\Big)\delta(\epsilon-\epsilon_{\bar{n},{\bf p+q}}-(-1)^n\omega)
\Bigg]\Big\rangle, \end{eqnarray}
 where $p=\sqrt{2m(\epsilon -\varepsilon_n)},~~n=1,2,$ $c=16n_se^2z_{12}/(9m\sigma_0),\ \delta=\Delta-\omega$ is the resonance detuning.
After angular integration we find
\begin{eqnarray} \label{s1a1}
&&\left(\begin{array}{c}
    s_1\\
    a_1
  \end{array}\right)
 =\mp \frac{e^{-\beta w}}{w+1}
  (e^{-\beta/4}-e^{-\beta})\int_0^\infty \frac{dy}{y}R_3(y)\times  \nonumber\\ &&\Bigg[R_1(y)\Big((3-4y^2)\frac{\vartheta(64(w+1)y^2-(3-4y^2)^2)}
{\sqrt{64(w+1)y^2-(3-4y^2)^2}}e^{-3\beta/4}\nonumber \\ && \pm (3+4y^2)\frac{\vartheta(64(w+1)y^2-(3+4y^2)^2)}
{\sqrt{64(w+1)y^2-(3+4y^2)^2}}\Big)- \nonumber \\ &&\Big(\pm R_1(y)+R_2(y)\Big)y\frac{\vartheta(4(w+1)-y^2)}
{\sqrt{4(w+1)-y^2}}\Big)e^{-3\beta/4}\Bigg],
\end{eqnarray}
\begin{eqnarray}
\label{s2a2} && \left(\begin{array}{c}
    s_2\\
    a_2
  \end{array}\right)=-\frac{e^{-\beta w}}{w+1/4}(e^{-\beta/4}-e^{-\beta})
  \int_0^\infty \frac{dy}{y}R_3(y)\times \nonumber \\ &&\Bigg[R_2(y)(3-4y^2)\frac{\vartheta(64(w+1/4)y^2-(3-4y^2)^2)}
{\sqrt{64(w+1/4)y^2-(3-4y^2)^2}}e^{-3\beta/4}\nonumber +\\&&\Big(R_1(y) \pm R_2(y)\Big)y\frac{\vartheta(4w+1-y^2)}
{\sqrt{4w+1-y^2}}\Bigg],
\end{eqnarray}
 where
$ w=2ma_B^2\epsilon,
 ~\vartheta(t)$ is the Heaviside function,  $R_1(y)=(\bar{V}_q)_{11},~~ R_2(y)=(\bar{V}_q)_{22},~~ R_3(y)= (\bar{V}_q)_{12},~~~ \beta=(2ma_B^{2}T)^{-1}, ~ y=qa_B$.

For  relaxation times, Eqs.(\ref{nuintra},\ref{nuinter}) yield:
 \begin{eqnarray} \label{tau1}
   &&\frac{1}{ \tau_1}=\frac{1}{\tau_0}\int_0^\infty \frac{dy}{y} \Bigg[(R_1(y))^2\frac{y\vartheta(4(w+1)-y^2)}
{(w+1)\sqrt{4(w+1)-y^2}} +\nonumber\\&& (R_3(y))^2\frac{8\vartheta(64(w+1)y^2-(3+4y^2)^2)}
{\sqrt{64(w+1)y^2-((3+4y^2))^2}} \Bigg],
\\
\label{tau12} &&\frac{1}{ \tau_{1,2}} =\frac{1}{\tau_0}
\int_0^\infty \frac{dy}{y} (R_3(y))^2\Big[1-\frac{(3+4y^2}{8(w+1)} \Big]\times \nonumber \\ &&\frac{8\vartheta(64(w+1)y^2-(3+4y^2)^2)}
{\sqrt{64(w+1)y^2-((3+4y^2))^2}},  \end{eqnarray}
\begin{eqnarray} \label{tau2}
   &&\frac{1}{ \tau_2}=\frac{1}{\tau_0}\int_0^\infty \frac{dy}{y} \frac{R_2^2(y)}{w+1/4}\Bigg[\frac{y\vartheta(4(w+1/4)-y^2)}
{\sqrt{4(w+1/4)-y^2}} +\nonumber\\&& R_3^2(y)\frac{8\vartheta(64(w+1/4)y^2-(3-4y^2)^2)}
{\sqrt{64(w+1/4)y^2-((3-4y^2))^2}} \Bigg],
\\
\label{tau21} &&\frac{1}{ \tau_{2,1}} =\frac{1}{\tau_0}
\int_0^\infty \frac{dy}{y} R_3^2(y)\Big[1+\frac{3-4y^2}{8(w+1/4)} \Big]\times \nonumber \\ &&\frac{8\vartheta(64(w+1/4)y^2-(3-4y^2)^2)}
{\sqrt{64(w+1/4)y^2-((3-4y^2))^2}},  \end{eqnarray}
  where $\tau_0=\pi ma_B^4\sigma_0/T.$ For functions $R_k(y) ~~(k=1,2,3)$  we have
\begin{eqnarray}
   R_1(y)=\frac{2y^2\Big[(y^2-4)^{1/2}-2\arccos(2/y)\Big]}{(y^2-4)^{3/2}}, \nonumber  \\
   R_2(y)=y^2\times \nonumber \\  \nonumber\frac{\Big[(y^2-1)^{1/2}(7+6y^2+2y^4)-(4+7y^2+4y^4)\arccos(1/y)\Big]}{8(y^2-1)^{7/2}},  \\
 \nonumber  R_3(y)=  \frac{16y^2\Big[(4y^2-9)^{1/2}(9+8y^2)-36y^2\arccos(3/(2y ))\Big]}{9\sqrt{2}(4y^2-9)^{5/2}}.
   \end{eqnarray}

Finally,  from Eqs.(\ref{j1}-\ref{an}), one can write the expressions for photogalvanic coefficients:

\begin{eqnarray}\label{33}
 \left(\begin{array}{c}
   \alpha_s \\
    \alpha_a
  \end{array}\right)
=-b\left(\begin{array}{c}
  f_s(\delta)~\Lambda_ s\\
   f_a(\delta)~\Lambda_a
 \end{array}\right),~~~~~b=\frac{2^7\sqrt{2}n_sme^9}{3^6\pi\sigma_0\kappa^3\hbar^5},
\end{eqnarray}
 \begin{eqnarray} \label{als}
 &&\Lambda_s=\int\limits_{-1/4}^\infty \frac{dw}{\tau_{1,2} \tau_{2,1}- \tau_1 \tau_2}\Bigg((w+1)
\tau_1 \tau_{2,1} \times\nonumber\\&& \Big[s_1 \tau_{1,2}+ s_2\tau_2\Big]+(w+\frac{1}{4})
\tau_2 \tau_{1,2}
  \Big[s_2 \tau_{2,1}+ s_1\tau_1\Big]\Bigg)+\nonumber\\&& \int\limits_{-1}^{-1/4} dw(w+1)
s_1 \tau_1 ,
\\ \label{ala} &&\Lambda_a=\int\limits_{-1/4}^\infty \frac{dw}{\tau_{1,2} \tau_{2,1}- \tau_1 \tau_2}\Bigg((w+1)
\tau_1 \tau_{2,1} \times \nonumber\\&&\Big[a_1 \tau_{1,2}+ a_2\tau_2\Big]+ (w+\frac{1}{4})
\tau_2 \tau_{1,2}
  \Big[a_2 \tau_{2,1}+ a_1\tau_1\Big]\Bigg)\nonumber\\&&+\int\limits_{-1}^{-1/4} dw(w+1)
a_1 \tau_1. \end{eqnarray}
We  restored $\hbar$ in (\ref{33}).
Functions $f_{s,a}(\delta)$ give the photogalvanic coefficients frequency dependence. Detuning $\delta$ and  width $\eta$ are measured in the inverse time, parameters $\Lambda_{s,a}$ have dimensionality of time. Parameter $b$ for He${}^4$ has a numerical value of $b=-1.486\times 10^{23}$cm${}^{3/2}$g${}^{-1/2}$s${}^{-2}$.

Eqs. (\ref{33}-\ref{ala}) give the main result of the paper. Naturally, infinitely small quantity $\eta$ should be replaced by a finite scattering rate (see below).

\subsection*{Discussion  and conclusions}
According to Eqs. (\ref{33}), the photogalvanic effect resonantly depends on resonance detuning $\delta$. PGE coefficient $\alpha_s$ has a symmetric   resonance and  $\alpha_a$ has an antisymmetric resonance as a function of $\delta$. The maximum value of $f_s(\delta)$ at $\delta=0$ is twice as big as the maximum of $f_a(\delta)$ at $\delta=\eta$. This behavior is similar to the PGE in semiconductor quantum wells. The  resonance originates from the intermediate state of indirect transitions due to the parallelism of subbands, rather than the resonance of direct intersubband optical transitions caused by conservation laws in the final states.

The finite value of $\eta$ is produced by the same scattering processes as these giving rise to the PGE. According to Eqs.(\ref{als},\ref{ala}) and Eqs.(\ref{s1a1},\ref{s2a2}), two characteristic groups of electrons are involved in the PGE, namely, the thermal electrons in the first subband with  $\epsilon\sim T$  and the photoexcited electrons with  $\epsilon-\Delta\sim T$. The finite subband width is determined by the intermediate optical excitation process. Hence, the reasonable approximation for $\eta$ is the  sum of their scattering rates $\eta\approx <<1/\tau_1>>+<<1/\tau_2>>$ (in the last expression the averaging denoted by $<<...>>$ should be done with the thermal distribution near the bands of the first and the second subbands, correspondingly).

The dependence of $\tau_{1}$ and $\tau_{2}$ on the energy at the conditions of the experiment \cite{alex_chep}, $n_S=1.7\times 10^6$ cm${}^{-2}$,  $T=0.1K$, is represented  in Fig.2. The scattering   becomes essentially weaker near threshold energies. Intersubband transition rates  $1/\tau_{1,2},~~1/\tau_{2,1}$ prove to be very weak, as compared with intrasubband  rates $1/\tau_1,~~1/\tau_2$.
\begin{figure}[h]\label{fig2}
\centerline{\epsfxsize=8cm\epsfbox{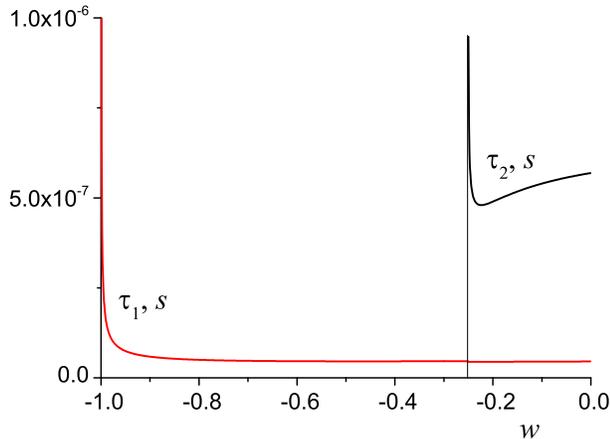}}
 \caption{Mean free times $\tau_1$ and $\tau_2$ {\it versus} the electron energy at $T=1$K. The vertical line marks the position of the second subband.}
\end{figure}
\begin{figure}[h]\label{fig3}
\centerline{\epsfxsize=7cm\epsfbox{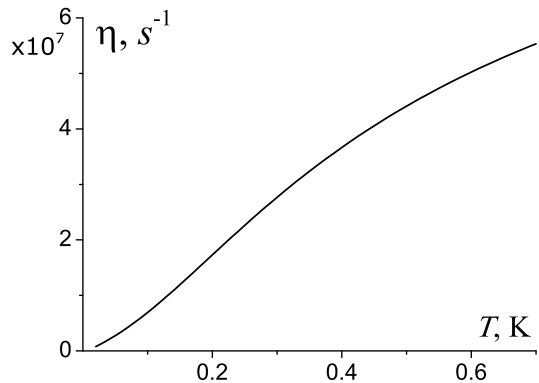}}
 \caption{Temperature dependence  of the resonance width $\eta$.}
\end{figure}
  The temperature dependence of photogalvanic coefficients is depicted in Fig.4.
 \begin{figure}[h]\label{fig4}
\centerline{\epsfxsize=6.5cm\epsfbox{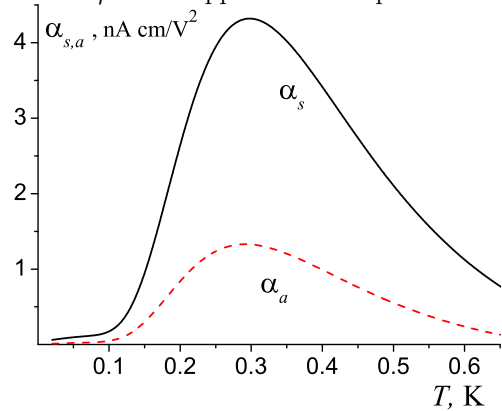}}
 \caption{Temperature dependence of frequency maxima of  PGE coefficients $\alpha_s$  at $\delta=0$ and $\alpha_a$ at $\delta=\eta$.}
\end{figure}
The strength of PGE is determined by $\alpha_s$ and $\alpha_a$. At $T=0.1K$ $\alpha_s(\delta=0)=1.57\times10^{-10}$ Acm/V${}^2$. This value is in reasonable accordance with the experimental value \cite{alex_chep} (obtained, however, in the presence of magnetic field). The PGE coefficients achieve the maximal values at temperature $T\approx 0.3$K:  $\alpha_s(\delta=0)=4.33\times10^{-9}$ Acm/V${}^2$,  $\alpha_a(\delta=\eta)=1.33\times10^{-9}$ Acm/V${}^2$ which are essentially larger  than the corresponding values for $T\sim 0.1$K. The ratio of linear to circular PGE coefficients maxima is 3.2.

It should be emphasized that large momentum transfer to ripplons in optical  transition amplitude $1\to 2$ enhances the transition process (and PGE coefficients). At the same time, the strong scattering increases the width of  resonance $\eta$ that suppresses the optical transitions. The photogalvanic coefficients extrema in Fig. 4 originate from the concurrence of these factors.

In conclusion, we have found the value of photocurrent along the charged helium surface  caused by tilted alternating electric field. The current contains two components representing responses to linear and circular polarization. The linear photogalvanic current has delta-like resonance and the circular photocurrent has antisymmetric resonance near the intersubband transitions frequency. The ripplon scattering  mechanism was taken into account. The current value  is consistent with that observed in the experiment.
\subsection*{Acknowledgements}
Authors are grateful to A. Chepelyanskii for attraction our attention to the paper \cite{alex_chep}.  This research was supported  by RFBR grant nos. 11-02-00060, 11-02-00730.

 \end{document}